\documentclass[aps,pra,superscriptaddress,longbibliography,twocolumn]{revtex4-2}
\usepackage{graphicx,amsmath,amssymb,mathtools,physics,enumitem,microtype}
\usepackage[dvipsnames]{xcolor}
\usepackage[utf8]{inputenc}
\usepackage[english]{babel}
\usepackage[colorlinks=true, linkcolor=MidnightBlue, citecolor=MidnightBlue, urlcolor=MidnightBlue]{hyperref}

\newcommand{\me}{\mathrm{e}}

\newcommand{\so}{\hat\sigma}

\newcommand{\So}{\hat{S}}

\newcommand{\App}[1]{App.~\ref{#1}}
\binoppenalty=\maxdimen
\relpenalty=\maxdimen

\newcommand{\tuda}{\affiliation{Institute for Applied Physics, TU Darmstadt, Hochschulstraße 4A, D-64289 Darmstadt, Germany}}
\newcommand{\fau}{\affiliation{Department of Physics, Friedrich-Alexander-Universität Erlangen-Nürnberg, Staudtstraße 7, D-91058 Erlangen, Germany}}
\newcommand{\mpl}{\affiliation{Max Planck Institute for the Science of Light, Staudtstraße 2, D-91058 Erlangen, Germany}}
\newcommand{\cesq}{\affiliation{CESQ/ISIS (UMR 7006), CNRS and Universit\'e de Strasbourg, 67000 Strasbourg, France}}

\begin{document}

\title{Scaling theory of decoherence in Dicke superradiance}
\author{Nico S. Bassler}
\email{nico.bassler@physik.tu-darmstadt.de}
\tuda
\author{Julian Lyne}
\fau
\mpl
\author{Javier Cuerda}
\email{javier.cuerda@physik.tu-darmstadt.de}
\tuda
\cesq

\begin{abstract}
The survival of many-body coherence depends on the competition between correlation buildup and decoherence. In Dicke superradiance, collective emission builds up correlations, producing a peak intensity scaling as $N^2$ for $N$ emitters. We develop a scaling theory including local dephasing and spontaneous emission and obtain fully collective, partially collective, and independent-emitter scaling regimes. The boundary of the fully collective regime defines a continuous phase transition in a transient observable. Local decoherence can prevent $N^2$ scaling despite increasing $N$.
\end{abstract}

\maketitle

\section{Introduction}\label{sec:introduction}
Achieving robust many-body coherence is a central challenge for modern platforms~\cite{streltsov2017colloquium,ladd2010quantum,burkard2023semiconductor,onizhuk2025colloquium,siddiqi2021engineering,kjaergaard2020superconducting,bruzewicz2019trapped,saffman2019quantum,gross2017quantum,altman2021quantum,browaeys2020manybody,amico2021roadmap,shermet2023waveguide,fauseweh2024quantum,daggett2026many} and emerging quantum technologies~\cite{acn2018the,pezze2018quantum,colombo2022entanglement,huang2024entanglement}. However, the inevitable coupling of a system to its environment results in decoherence, which can destroy coherence between particles or restrict its buildup~\cite{zurek2003decoherence,yu2004finite,carvalho2004decoherence,aolita2008scaling,yu2009sudden,wang2010sudden,levi2016robustness,suter2016colloquium,kirton2017suppressing,deleon2021materials,mostafa2026signatures}. Therefore, the competition between coherent processes and local decoherence, and its resolution in the thermodynamic limit, are central problems for open quantum many-body systems~\cite{daley2014quantum,aolita2015open,sieberer2016keldysh,minganti2018spectral,mori2024liouvillian,Fazio2025,sieberer2025universality}.

\begin{figure}[!ht]
    \centering
    \includegraphics[width=0.95\linewidth]{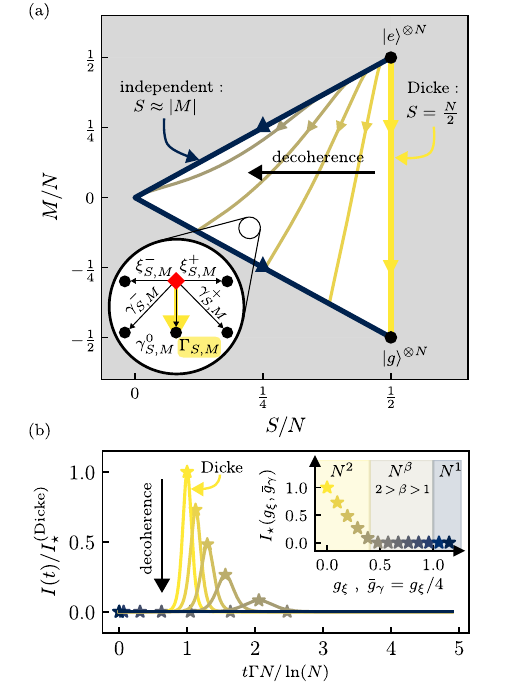}
    \caption{\textbf{Dicke triangle, burst scalings and transitions. }{(a)} Dynamics on the Dicke triangle, where for increasing decoherence strength trajectories are pushed towards smaller $S$--values. The inset shows the matrix elements of the local and collective processes for a given state $\ket{S,M}$ (\rotatebox{45}{\textcolor{red}{\rule{2mm}{2mm}}}). 
    (b) Intensity dynamics for the trajectories in (a), normalized by the peak intensity $I_\star^{\mathrm{(Dicke)}}$ of ideal Dicke superradiance at the same $N$. Decoherence reduces the ideal superradiant burst. The inset shows the peak intensity as a function of the decoherence strength undergoing a phase transition.}
    \label{fig:fig1}
\end{figure}

Rather than generating many-body coherence through unitary Hamiltonian dynamics, structured dissipation can drive a system toward correlated states~\cite{diehl2008preparation,diehl2008quantum,Verstraete2009,harrington2022engineered,zhan2026rapid}. A minimal example of the latter is Dicke superradiance, where collective emission of $N$ quantum two-level emitters with rate $\Gamma$ dynamically generates many-emitter coherence~\cite{dicke1954}. Starting from full initial excitation, the system evolves through the highly correlated fully symmetric Dicke states~[see Fig.~\ref{fig:fig1}(a)]. The dynamically generated coherence manifests itself directly in the radiated intensity \(I(t)=\Gamma\sum_{i,j}\langle\hat\sigma_i^\dagger\hat\sigma_j\rangle\), where \(\hat\sigma_j\) is the two-level lowering operator for particle $j$ [see Fig.~\ref{fig:fig1}(b)]. The intensity takes the form of a short, intense burst at time $t_\star \approx \ln N/(N\Gamma)$ with quadratic peak intensity $I_\star \propto N^2$~\cite{dicke1954,gross_haroche}. Dicke superradiance has been studied in depth theoretically~\cite{gross_haroche,rehler1970superradiance,agarwal1970master,bonifacio1971quantum1,bonifacio1971quantum2,degiorgio1971approximate,haake1972quantum,lee1976transition,lee1977exact1,lee1977exact2,lemberger2021radiation,robicheaux2021theoretical,malz2022,masson2022universality,cardenas2023many,lohof2023signatures,masson2024dicke,Mok2025,holzinger2025solving,holzinger2025compact,windt2025effcts,Freter2025,bassler2025absence,holzinger2025superradiant,holzinger2025symbolic,rosario2025unraveling,zhang2025unraveling,rahimi2025timing,lyne2026dicke,alabbar2026initiation,rusconi2026optical,belliardo2026extracting,winter2026extensive,prazeres2026kinetically} and experimentally demonstrated across a diverse range of platforms~\cite{skribanowitz1973observation,gross1976observation,gibbs1977single,devoe1996observation,wang2007superradiance,scheibner2007superradiance,rohlsberger2010collective,mlynek2014observation,goban2015superradiance,cong2016Dicke,solano2017super,bradac2017room,okaba2019superradiance,adl2023superradiance,pallmann2024cavity,ferioli2025emergence,bach2026emergence,wei2026chiral,douglas2026many,du2026programmable}.

In realistic systems, coherence buildup competes with local decoherence of individual emitters, which generally reduces coherence and degrades the ideal superradiant burst~\cite{maki1989influence,temnov2005superradiance,shammah2017superradiance,shammah2018open,suarez2022superradiance,stryzhenko2024N,sun2025engineering}, as illustrated in Fig.~\ref{fig:fig1}. If the effective collective decay rate scales linearly with the system size (extensively), local decoherence processes that do not scale with the system size (intensively) are negligible in the thermodynamic limit ($N\rightarrow\infty$)~\cite{malz2022}. However, the thermodynamic limit need not follow this scaling behavior and the relative scaling of local and collective processes will generally depend on the experimental platform and the manner in which the system size is scaled up. This raises the question of which system-size-dependent scalings of the relevant processes lead to fully collective emission in the thermodynamic limit and which do not.

Here, we introduce an analytical scaling theory that clarifies the role of local decoherence in Dicke superradiance. Considering local spontaneous emission with rate $\gamma$ and local dephasing with rate $\xi$, we determine how these processes must scale relative to the collective emission rate to preserve or suppress fully collective emission. In particular, we study the dynamics governed by the Lindblad master equation
\begin{equation}\label{eq:master_intro}
\dot\rho = \Gamma\,\mathcal{D}[\So]\rho + \xi \sum_{j=1}^{N}\mathcal{D}[\hat\sigma^z_j]\rho + \gamma\sum_{j=1}^{N}\mathcal{D}[\so_j]\rho ,
\end{equation}
where $\mathcal{D}[\hat X]\rho = \hat X \rho \hat X^\dag - \tfrac{1}{2}(\hat X^\dag \hat X \rho + \rho \hat X^\dag \hat X)$, $\So=\sum_{j=1}^N \so_j$ and $\hat\sigma^z_j= (\hat{\sigma}^\dag_j\hat{\sigma}^{\phantom\dag}_j-\hat{\sigma}^{\phantom\dag}_j\hat{\sigma}^\dag_j)/2$. Permutational symmetry enables a reduced description in the Dicke basis [see~Fig.~\ref{fig:fig1}(a)], which has been derived and utilized in previous works \cite{lee1975diagrammatic,chase2008collective,Baragiola2010,xu2013simulating,gegg2016efficient,shammah2017superradiance,kirton2017suppressing,shammah2018open,Zhang2018,Bastin2025}. The resulting rate equations describe the flow of population between the states \(\ket{S,M}\), with $S=0,1,\ldots N/2$ (assuming $N$ is even) and $M=-S,-S+1,\ldots, +S$, referred to as the Dicke triangle~[see Fig.~\ref{fig:fig1}(a)]. The matrix elements of collective emission $\hat S\ket{S,M}=\sqrt{(S+M)(S-M+1)}\ket{S,M-1}$ preserve the collective spin $S$ and decrease in magnitude as $S$ is decreased. The local processes do not conserve the collective spin and push the trajectories into the interior of the Dicke triangle, as indicated by the inset in Fig.~\ref{fig:fig1}(a) [see \App{app:pi}]. The model in Eq.~\eqref{eq:master_intro} isolates the competition of local and collective dynamics without any unitary dynamics, allowing for a reduction to mean-field-like equations of motion in the large-$N$ limit, enabling an analytical solution to the asymptotic dynamics [see \App{app:large_N} for details].

We study the scaling of the peak emission intensity $I_\star$, which captures the buildup of correlations in the presence of local decoherence, revealing the generalized scaling law 
\begin{equation}\label{eq:scaling_law_intro}
I_\star \propto N^{\beta(g_\xi,\bar g_\gamma)}\quad \text{with}\quad 1 \leq \beta(g_\xi,\bar g_\gamma) \leq 2\;,
\end{equation}
where the analytical expression for the scaling exponent $\beta(g_\xi,\bar g_\gamma)$ given in Eq.~\eqref{eq:beta_general} is a function of the dimensionless scaling variables
\begin{equation}\label{eq:scaled_params_intro}
    g_\xi = \frac{\xi}{N\Gamma}\,, \qquad \bar g_\gamma = \frac{\gamma \ln N}{N\Gamma}\,.
\end{equation}
This indicates that local decoherence may not only prevent a quadratic scaling, but can continuously tune the peak intensity between the fully collective and independent limits, defining three regimes [see inset in Fig.~\ref{fig:fig1}(b)]. The value $\beta=2$ separates the fully collective regime with quadratic peak intensity from a partially collective regime with subquadratic peak intensity ($2>\beta>1$), and the value $\beta=1$ separates the partially collective regime from the linear independent-emitter regime. In the thermodynamic limit $N\rightarrow\infty$, the peak intensity at the fully collective boundary ($\beta=2$) becomes non-analytic, resembling a continuous phase transition. In contrast to dissipative phase transitions that occur in the steady state~\cite{morrison2008dynamical,kessler2012dissipative,carmichael2015breakdown,sieberer2016keldysh,dallatorre2016dicke,kirton2017suppressing,kirton2018superradiant,kirton2019introduction,minganti2018spectral,shimshi2024quantum,leppenen2024quantum,sieberer2025universality}, the transition ($\beta=2$) identified here occurs in a transient observable, while the steady-state remains trivial with all emitters in the ground state. For simplicity, we use phase transition terminology to characterize the obtained transition, but stress that it does not fall within the standard paradigm. Recently, Ref.~\cite{windt2026quantum} explored bosonic superradiance with additional repulsive interactions, also discovering a generalized scaling law for the peak intensity with a phase transition between quadratic and subquadratic scaling.

\begin{figure*}[!ht]
    \centering
    \includegraphics[width=0.95\linewidth]{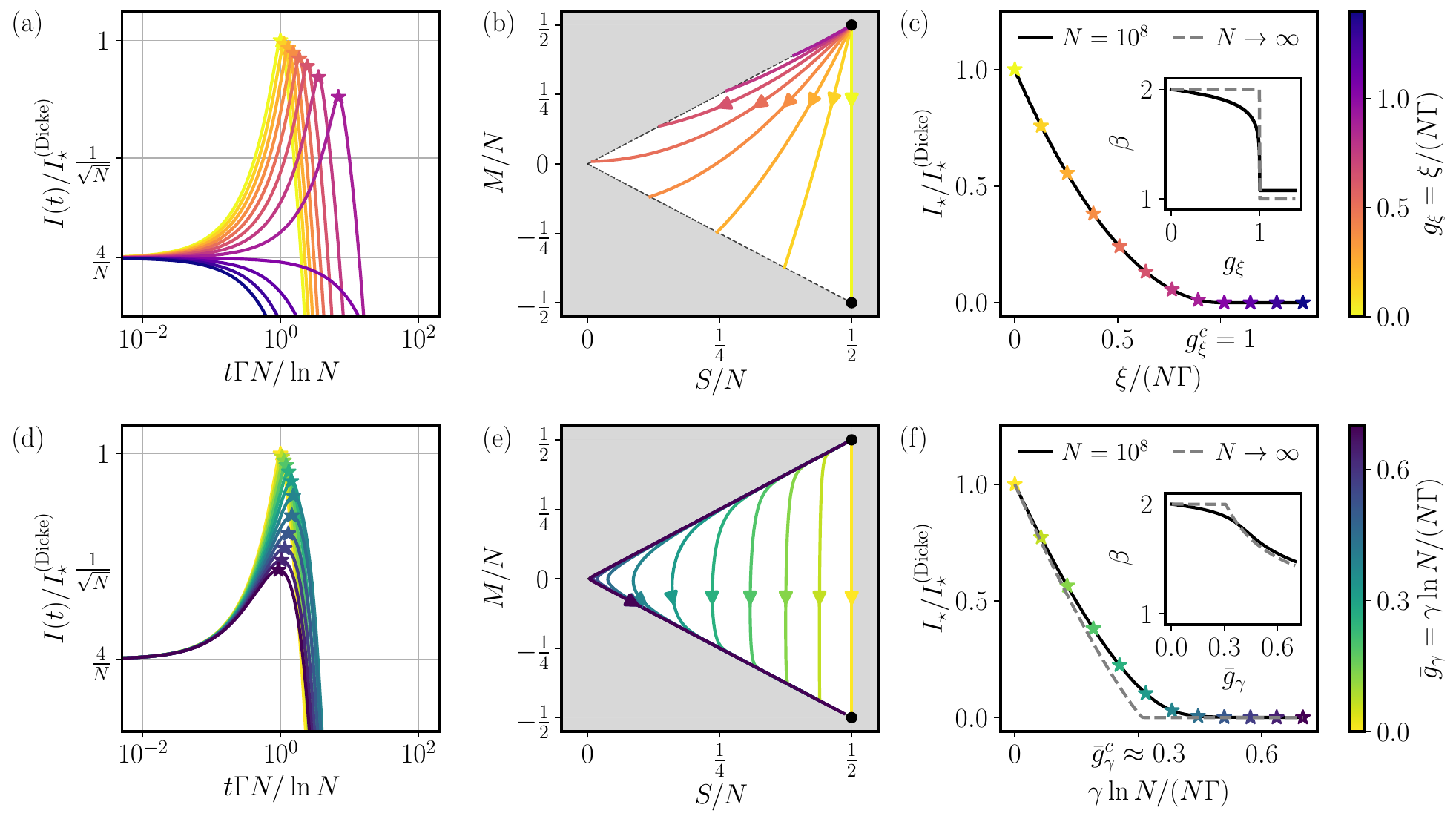}
    \caption{\textbf{Large--$N$ dynamics.} Results are shown for dephasing (top row) and spontaneous emission (bottom row) at $N = 10^8$. 
(a,d) Normalized intensity as a function of rescaled time for varying scaling variables on a log-log scale. Stars ($\star$) indicate the peak intensity $I_\star$.
(b,e) Corresponding trajectories on the Dicke triangle. While dephasing changes the angle at which the trajectories enter the triangle, local spontaneous emission shifts the point at which the trajectories enter the triangle to smaller $S$--values.
(c,f) Normalized peak intensity undergoing a phase transition as a function of the scaling variables $g_\xi,\bar g_\gamma$ and insets showing the scaling exponent $\beta$.
For dephasing (c), $I_\star$ closes quadratically at the critical value $g_\xi^c = 1$ and $\beta$ changes in a discontinuous manner, showing a transition from the fully collective to the independent-emitter regime. In contrast, for spontaneous emission (f), $I_\star$ closes linearly at the critical value $\bar g_\gamma^c = 1 - \ln 2 \approx 0.3$ and $\beta$ decreases continuously, showing a transition from the fully collective to the partially collective regime.}
    \label{fig:fig2}
\end{figure*}

\section{Local Dephasing}

We first consider only collective emission and local dephasing without local spontaneous emission (\(\gamma=0\)), which provides a controlled reference problem for the general case. In this case, the large-\(N\) dynamics are analytically solvable and are governed solely by the scaling variable \(g_\xi\) [see Eq.~\eqref{eq:scaled_params_intro}]. We define the intensive variables $\mathfrak{s}=S/N$, $\mathfrak{m}=M/N$, and the normalized intensity of the superradiant channel used throughout the manuscript
\begin{equation}\label{eq:intensiveI}
\mathfrak{I}=\frac{\langle \hat S^\dag \hat S \rangle}{N^2}=\frac{I}{\Gamma N^2}\approx \mathfrak{s}^2-\mathfrak{m}^2,
\end{equation}
which remains finite as $N\to\infty$ in the ideal Dicke scenario. We expand the Dicke-basis transition rates to lowest order in $1/N$, and make the hydrodynamic large-$N$ approximation, reducing the exact stochastic dynamics at finite $N$ to deterministic equations for the intensive variables~\cite{kurtz2009,malz2022,Mok2025} [see \App{app:large_N}]. We refer to them as mean-field equations because they describe the leading large-\(N\) dynamics of the intensive collective variables, but the approximation does not rely on a microscopic operator-factorization. With the rescaled time $\tau=N\Gamma t$, the equations of motion take the form
\begin{subequations}\label{eq:mean-field-dephasing}
\begin{align}
    \partial_\tau\mathfrak{I}(\tau)&= \left[2\mathfrak{m}(\tau) - g_\xi\right]\mathfrak{I}(\tau),
    \label{eq:mean-field-dephasing-I}\\
    \partial_\tau\mathfrak{m}(\tau)&= -\mathfrak{I}(\tau).
    \label{eq:mean-field-dephasing-m}
\end{align}
\end{subequations}

The deterministic equations are independent of \(N\), and the system-size dependence enters through the initial condition \(\mathfrak I(\tau=0)=1/N\), which initiates the subsequent amplification of the intensity. This value is the exact finite-\(N\) normalized collective intensity of the fully inverted state and is hereafter referred to as the seed [see \App{app:mean_field_justification}]. The resulting asymptotic scaling is verified against exact stochastic simulations in Fig.~\ref{fig:figS1} and Apps.~\ref{app:finite_size} and \ref{app:finite_size_spontaneous}.

While the coupled nonlinear differential equations in Eqs.~\eqref{eq:mean-field-dephasing} are analytically solvable [see \App{App:sect-analytical-mf}], the qualitative behavior can already be understood from the behavior at $\tau =0$. The factor $2\mathfrak m(\tau)-g_\xi$ in Eq.~\eqref{eq:mean-field-dephasing-I} acts as an instantaneous rate for the intensity, causing growth or decay from the seed value. A lower positive rate reduces and delays the superradiant peak as shown in Fig.~\ref{fig:fig2}(a), while a negative rate indicates the decay of the intensity and therefore the absence of a superradiant burst. With a fully excited initial state ($\mathfrak m(0)=1/2$) one finds \(\partial_\tau \mathfrak I(0)=(1-g_\xi)\mathfrak I(0)\). The initial amplification factor \(1-g_\xi\) determines whether the seed is amplified and fixes the initial direction in which the trajectory leaves the fully excited state, as visible in Fig.~\ref{fig:fig2}(b), fixing its subsequent evolution. While collective emission causes a fully excited $N$-emitter system to move downward on the Dicke triangle, dephasing pushes the dynamics towards the independent edge $S\approx\abs{M}$ of the Dicke triangle, as also pointed out in Ref.~\cite{shammah2017superradiance}. For $g_\xi<1$, the initial flow points towards the bulk of the Dicke triangle, where $S\propto N$, such that the peak intensity of the burst then scales as $I_\star\propto N^2$. For $g_\xi\ge1$, the trajectory is directed towards the independent-emitter edge $S\approx\abs{M}$, and no amplification occurs, resulting in linear peak intensity scaling $I_\star\propto N$.

From the analytical solutions of Eqs.~\eqref{eq:mean-field-dephasing}, we obtain the mean-field peak intensity and peak time [see \App{App:sect-analytical-mf}]
\begin{equation}
\mathfrak{I}_{\star}=(1-g_\xi)^2/4,\qquad \tau_\star\approx \ln N/(1-g_\xi),
\end{equation}
in agreement with the exact Monte Carlo simulations [see \App{app:finite_size}]. The normalized peak intensity $I_\star/I_\star^{\rm(Dicke)}=4\mathfrak I_\star$ acts as an order parameter for the fully collective-to-independent transition, where, within mean-field $I_\star^{\rm(Dicke)}=\Gamma N^2/4$. This order parameter is unity in the ideal Dicke limit and vanishes for $g_\xi \ge 1$~[see Fig.~\ref{fig:fig2}(c)]. Since the peak intensity closes quadratically as $g_\xi\to1^{-}$, the function $\mathfrak{I}_{\star}(g_\xi)$ is non-analytic in its second derivative at the critical point $g_\xi^c = 1$, resembling a continuous phase transition with critical exponent $\nu=2$. This non-analyticity is already present in the mean-field result at finite $N$, and the exact dynamics recover it only in the thermodynamic limit $N\to\infty$ [see \App{app:finite_size}].

\section{Local Spontaneous Emission}\label{sec:spontaneous_emission}

We now include local spontaneous emission in addition to local dephasing and collective emission. Performing a mean-field analysis in the large-$N$ limit [see \App{app:large_N}], we obtain the equations of motion
\begin{subequations}
\label{eq:mf_complete}
\begin{align}
    \partial_\tau\mathfrak{I}(\tau) &= \left[ 2\mathfrak{m}(\tau) - g_\xi-g_\gamma\right] \mathfrak{I}(\tau),
    \label{eq:mf_complete_I}\\
    \partial_\tau\mathfrak{m}(\tau) &= -\mathfrak{I}(\tau) -g_\gamma\left[\mathfrak{m}(\tau) + \tfrac{1}{2}\right],
    \label{eq:mf_complete_m}
\end{align}
\end{subequations}
where $g_\gamma = \gamma/(N\Gamma)$. We show the dynamics without dephasing (\(g_\xi=0\)) of the intensity in Fig.~\ref{fig:fig2}(d), trajectories on the Dicke triangle in Fig.~\ref{fig:fig2}(e), and the corresponding peak intensity in Fig.~\ref{fig:fig2}(f). The reduction of the peak intensity without dephasing is facilitated by the depletion of excitations by spontaneous emission~[see Eq.~\eqref{eq:mf_complete_m}] and the accompanying destruction of coherence. Note that the intensity includes only emission through the collective channel.

Unlike the dephasing-only case, Eqs.~\eqref{eq:mf_complete} do not admit a closed analytical solution. Following the initial dynamics analysis from the previous section, we identify the initial amplification condition for the intensity as \(g_\xi+g_\gamma<1\) [see Eq.~\eqref{eq:mf_complete_I}]. However, this condition is no longer sufficient to determine whether fully collective scaling is reached because spontaneous emission depletes the inversion while the intensity is amplified. Instead, one must determine whether enough excitations remain before the gain of the intensity vanishes. The competition between excitation loss and intensity amplification is determined during an early-time amplification stage, in which the intensity grows from its initial seed while spontaneous emission depletes the inversion. During the early-time amplification stage, the trajectory stays very close to the $\abs{M}=S$ boundary~[see Fig.~\ref{fig:fig2}(e)] since the collective channel does not yet appreciably deplete the inversion. Thus, during the amplification stage \(\mathfrak I(\tau)\ll g_\gamma(\mathfrak m(\tau)+1/2)\) and the equations of motion at early times can be approximated as [see \App{app:asymptotic_mean}]
\begin{subequations}
\begin{align}
    \label{eq:mf_complete_lin_I}
    \partial_\tau\mathfrak{I}^{\text{lin}}(\tau) &= \left[ 2\mathfrak{m}(\tau) - g_\xi-g_\gamma\right] \mathfrak{I}^{\text{lin}}(\tau),\\
    \label{eq:mf_complete_lin_m}
    \partial_\tau\mathfrak{m}(\tau) &\approx -g_\gamma\left[\mathfrak{m}(\tau)+\tfrac{1}{2}\right],
\end{align}
\end{subequations}
where there is no backaction of $\mathfrak I$ on $\mathfrak m$ and Eq.~\eqref{eq:mf_complete_lin_m} can be integrated to give \(\mathfrak m(\tau)=\me^{-g_\gamma\tau}-1/2\). During this time, the intensity experiences the amplification rate \(2\mathfrak{m}(\tau)-g_\xi-g_\gamma\). The peak of the linearized intensity occurs when the amplification rate vanishes, \(2\mathfrak m(\tau)=g_\xi+g_\gamma\). Thus, the total amplification, and hence the peak intensity, is set by the
integral of this decaying amplification rate 
\(\ln \mathfrak I_\star^{\mathrm{lin}}
=
    -\ln N
    +
    \int_0^{\tau_\star^{\mathrm{lin}}}
    \left(2\mathfrak m(\tau)-g_\xi-g_\gamma\right)\text{d}\tau ,
\)
up to that time, where \(2\mathfrak m(\tau_\star^{\mathrm{lin}})=g_\xi+g_\gamma\).

\begin{figure}
    \centering
    \includegraphics[width=\linewidth]{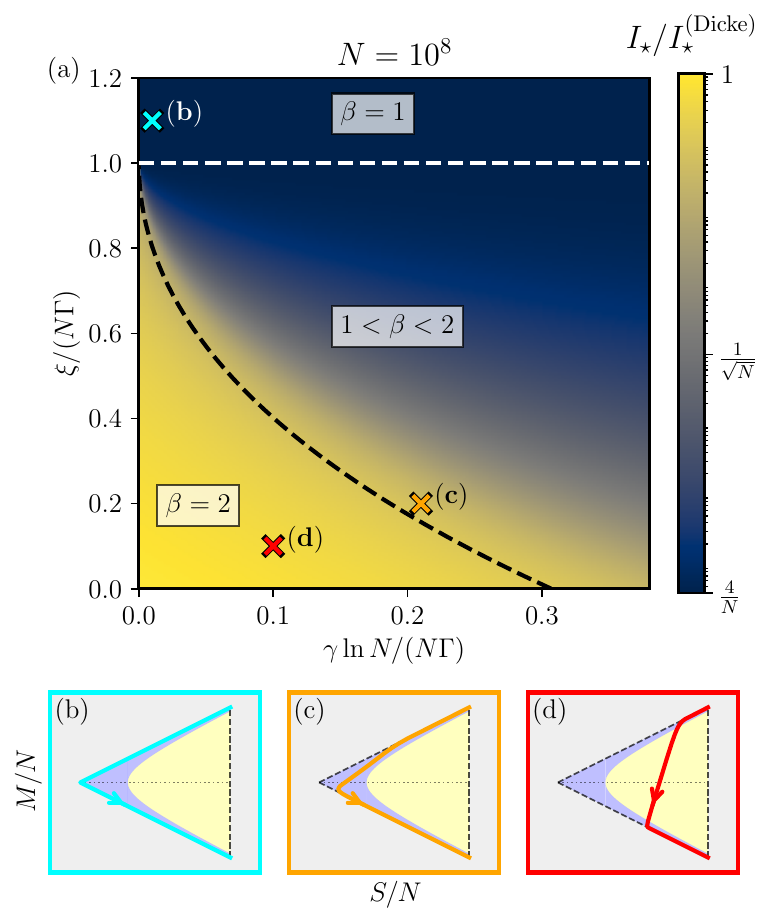}
    \caption{\textbf{Phase diagram and Dicke-triangle trajectories. }(a) Mean-field phase diagram showing the normalized intensity in the $(\bar g_\gamma, g_\xi)$ plane for $N=10^8$. The dashed lines indicate the asymptotic phase boundaries at $\beta=2$ (black) and $\beta=1$ (white). In (b-d) we show corresponding representative trajectories in the three regions. Blue shading in panels (b-d) marks the region $\mathfrak I=\mathfrak s^2-\mathfrak m^2<N^{-0.2}$, corresponding to $\beta<1.8$ and with the complement in yellow. The specific threshold is chosen only for visualization.}
    \label{fig:fig3}
\end{figure}

For the intensity to reach $\mathfrak{I}_\star= \mathcal{O}(1)$, corresponding to fully collective scaling $I_\star \propto N^2$, the initial seed $\mathfrak{I}(0) = 1/N$ must be amplified by a factor of order $N$ and must therefore last $\tau\propto\ln N$ (peak time). Spontaneous emission depletes the gain on a timescale $\tau \propto 1/g_\gamma$, and therefore suppresses fully collective scaling unless $g_\gamma \lesssim 1/\ln N$. This identifies $\bar g_\gamma = g_\gamma \ln N$ as the natural scaling variable, where a finite value of $\bar g_\gamma $ still allows for a scaling of the intensity as $N^2$. An alternative thermodynamic limit obtained by holding $g_\gamma$ fixed is discussed in \App{app:alternative_phasediagram}. Substituting \(g_\gamma=\bar g_\gamma/\ln N\), we find the intensity as $\mathfrak I_\star^{\mathrm{lin}}=N^{\beta(g_\xi,\bar g_\gamma)-2}$ with exponent [see \App{app:asymptotic_mean}]
\begin{equation}
\label{eq:beta_general}
\begin{aligned}
    \beta(g_\xi,\bar g_\gamma)&=1+\frac{1-g_\xi-(1+g_\xi)\ln\left(\frac{2}{1+g_\xi}\right)}{\bar g_\gamma},
\end{aligned}
\end{equation}
which is obtained in the limit \(g_\gamma=\bar g_\gamma/\ln N\to0\) at fixed \(\bar g_\gamma\). The expression in Eq.~\eqref{eq:beta_general} holds in the partially collective regime \(1<\beta<2\). Its intersections with \(\beta=2\) and \(\beta=1\) determine the boundaries to the fully collective and independent-emitter regimes, respectively. Beyond these boundaries, the exponent remains fixed at \(\beta=2\) and \(\beta=1\) [see Fig.~\ref{fig:figS1}]. Since the amplification rate changes sign at $2\mathfrak{m}(\tau) = g_\xi+g_\gamma$, the peak intensity for $\beta<2$ is reached entirely within the linearized regime. For $\beta=2$, the intensity reaches $\mathfrak{I}=\mathcal{O}(1)$ and the subsequent bulk evolution sets only the prefactor. In both cases, the scaling exponent is determined by the early-time amplification stage alone and $I_\star\propto N^{\beta(g_\xi,\bar g_\gamma)}$. 

From the peak intensity as a function of the scaling variables $g_\xi, \bar{g}_\gamma$, we obtain a phase diagram that is organized by the scaling exponent \(\beta\) [see Fig.~\ref{fig:fig3}(a)]. The collective boundary \(\beta=2\) is indicated by the black dashed line. For \(g_\xi=0\), this yields \(\bar g_\gamma^c=1-\ln 2\approx0.3\)~[see Fig.~\ref{fig:fig2}(f)]. Across this boundary lies a partially collective regime, with subquadratic peak intensity \(I_\star\propto N^\beta\) with \(1<\beta<2\). The representative trajectories in Fig.~\ref{fig:fig3}(c,d) illustrate this distinction: in the fully collective regime, the trajectory enters the bulk of the Dicke triangle, whereas in the partially collective regime the dynamics are restricted to a boundary layer along the $\abs{M} = S$ boundary [see \App{app:dicke_triangle_scaling}]. The second boundary, \(\beta=1\), separates this partially collective regime from independent-emitter scaling \(I_\star\propto N\), where the trajectory remains close to the boundary \(S\approx\abs{M}\) throughout [see Fig.~\ref{fig:fig3}(b)].

In the thermodynamic limit, the normalized peak intensity closes differently for $\bar g_\gamma\neq 0$ than in the dephasing-only case. Expanding around a point on the phase-boundary $\beta = 2$~[see \App{app:asymptotic_mean} for details], we find that the boundary with spontaneous emission has a linear closing ($\nu = 1$), in contrast to the quadratic closing ($\nu = 2$) along the dephasing-only axis [compare Figs.~\ref{fig:fig2}(c, f)].

\section{Conclusion}\label{sec:conclusion}

We analyzed the scaling behavior of the peak intensity in Dicke superradiance with local dephasing and local spontaneous emission as a diagnostic of dynamically generated many-body coherence in the presence of local decoherence. We identified the scaling variables \(g_\xi\) and \(\bar g_\gamma\), which define a nontrivial thermodynamic limit. The scaling variables are emergent quantities that arise from the competition of distinct dynamical processes and cannot be derived from an individual process alone. For the peak intensity, we derived the generalized scaling law \(I_\star\propto N^{\beta(g_\xi,\bar g_\gamma)}\), which organizes the dynamics into three regimes: a fully collective regime, \(I_\star\propto N^2\), a partially collective regime \(I_\star\propto N^\beta\) with \(1<\beta<2\), and the independent-emitter regime \(I_\star\propto N\). The boundary at \(\beta=2\) constitutes a continuous phase transition in a transient observable.

The scaling theory presented here can be applied to experiments by identifying how the rates \(\Gamma(N)\), \(\xi(N)\), and \(\gamma(N)\) scale with $N$. This generally determines a path in the (\(g_\xi,\bar g_\gamma\)) plane as the system size is varied, since different scaling behaviors do not preserve the ratios between collective emission and local decoherence rates. For example, in cavity-QED settings, where superradiance is facilitated by collective coupling to a lossy cavity mode, the light-matter coupling is commonly normalized as \(g_0/\sqrt N\)~\cite{hepp1973equilibrium,dimer2007proposed,gelhausen2018dissipative,kirton2019introduction,fiorelli2023mean}. This normalization is chosen such that single-particle rates and energies remain finite in the thermodynamic limit. It results in a collective decay rate $\Gamma \propto 1/N$ after adiabatic elimination of the cavity, and therefore the scaling variables for this normalization are $g_\xi=\text{const.}$ and $\bar g_\gamma\propto \ln N$. Consequently, increasing the emitter number $N$ logarithmically increases the influence of spontaneous emission, and therefore results in independent emission in the thermodynamic limit.

More generally, the asymptotic behavior of an open many-body system depends not only on the processes that generate and destroy correlations but also on how their rates scale relative to the timescale over which correlations build up. The distinct scaling variables identified here provide a concrete example of how this competition can produce different scaling regimes and critical behavior in a transient observable. Increasing \(N\) alone therefore does not guarantee quadratic \(N^2\) scaling in Dicke superradiance.

\section*{Acknowledgments}
We acknowledge fruitful discussions with A. Jaber, J. Dubail, J. Schachenmayer, and C. Genes. Numerical simulations were performed in Julia~\cite{Julia-2017}, using the package DifferentialEquations.jl~\cite{Rackauckas2017}, with figures produced using PyPlot.jl and Matplotlib~\cite{Hunter:2007}. We acknowledge financial support from the Max Planck Society and from the ERC Consolidator project MATHLOCCA (Grant No. 101170485). 

\section*{Data and Code Availability}
The complete numerical framework, including the exact permutationally invariant Lindblad solvers, mean-field integration scripts, and Monte Carlo simulation routines required to reproduce all main-text and appendix figures, is available on GitHub~\cite{SuperradianceMoleculesCode}.

\newpage

\clearpage
\appendix

\setcounter{figure}{0}
\renewcommand{\thefigure}{A\arabic{figure}}
\renewcommand{\theHfigure}{A\arabic{figure}}

\renewcommand{\topfraction}{0.99}
\renewcommand{\textfraction}{0.01}
\renewcommand{\floatpagefraction}{0.95}

\setcounter{topnumber}{3}
\setcounter{totalnumber}{5}

\renewcommand{\dbltopfraction}{0.99}
\renewcommand{\dblfloatpagefraction}{0.8}
\setcounter{dbltopnumber}{2}

\section{Permutationally invariant dynamical model}\label{app:pi}
In this section, we give further details of the full dynamical model described in the main text; see Eq.~\eqref{eq:master_intro}. The dynamics described by this model possess two important simplifications. First, the Lindbladian is invariant under permutations of the emitters.  Since the initial state $\ket{e,\ldots,e}$ is fully symmetric ($\ket{e}$ denoting the excited state), the density matrix remains permutationally invariant during the time evolution. This symmetry naturally suggests working in the Dicke basis of total spin eigenstates $\ket{S,M}$, where $S$ denotes the total spin quantum number and $M$ its projection onto the $z$--axis. Second, for the dissipators in Eq.~\eqref{eq:master_intro}, the population and coherence sectors decouple in the permutation-invariant basis, and no coherences are generated in the Dicke basis, reducing the dynamics to a Markov chain.

The total Hilbert space $\mathcal H_{\text{tot}}$ constructed as the tensor product over the Hilbert spaces of $N$ spin-$1/2$ particles $\mathcal H_{1/2}$ (two-level emitters with internal states $\ket{e},\ket{g}$) can be decomposed as 
\begin{equation}
\mathcal H_{\text{tot}}
=\bigotimes_{j=1}^{N}\mathcal H_{1/2}
\cong\bigoplus_{S=0}^{N/2}\bigoplus_{\alpha=1}^{d_S}\mathcal H^{(\alpha)}_S ,
\end{equation}
where the index $\alpha$ labels the $d_S$--fold degeneracy of the $2S+1$ dimensional subspace with spin $S$, but plays no role for permutation-invariant states and dynamics. 
Consequently, the density matrix can be fully characterized by the populations
\begin{equation}
    \begin{aligned}
        p_{S,M}&=\Tr\!\left[\rho\,\Pi_{S,M}\right],\\
        \Pi_{S,M}&=\sum_{\alpha=1}^{d_S}\outerproduct{S,M,\alpha}{S,M,\alpha}.
    \end{aligned}
\end{equation}

The time evolution of the system described by the master equation in Eq.~\eqref{eq:master_intro} therefore reduces to a set of linear differential equations for the populations $p_{S,M}$, where the transition rates for collective decay $\Gamma_{S,M}$, local dephasing $\xi^{\pm}_{S,M}$, and local spontaneous emission $\gamma^{\pm,0}_{S,M}$ are known functions of $(S,M,N)$ and are given below [see also Refs.~\cite{shammah2018open,chase2008collective,Baragiola2010,Zhang2018,Bastin2025}]. In practice, we do not simulate the rate equations, but the underlying Markov process using Gillespie Monte Carlo, where one stochastically evolves a state $(S,M)$ [see also Ref.~\cite{Zhang2018}], except for the large-scale phase-diagram calculations in Fig.~\ref{fig:figS1}(c), for which we use adaptive tau-leaping. For completeness, we restate the rates for the relevant local and collective processes in the Dicke basis~\cite{chase2008collective,Zhang2018,shammah2018open}

\begin{equation}
\label{eq:dephasing_rates}
    \begin{aligned}
    \Gamma_{S,M} &= \Gamma (S+M)(S-M+1),\\
    \xi^{+}_{S,M} &= \xi\frac{(S-M+1)(S+M+1)(N-2S)}{4(S+1)(2S+1)}, \\
    \xi^{-}_{S,M} & = \xi\frac{(S-M)(S+M)(N+2S + 2)}{4S(2S+1)},\\
    \gamma_{S,M}^{+}&=\gamma\frac{\left(S-M+1\right)\left(S-M+2\right)\left(N-2S\right)}{4\left(S+1\right)\left(2S+1\right)},\\
    \gamma_{S,M}^{0}&= \gamma\frac{\left(S+ M\right)\left(S- M+1\right)(2+N)}{4S\left(S+1\right)},\\
    \gamma_{S,M}^{-}&=\gamma\frac{\left(S+M\right)\left(S+M-1\right)\left(N+2S+2\right)}{4S\left(2S+1\right)}.
\end{aligned}
\end{equation}
The superscripts in the local dephasing rates $\xi^{\delta}_{S,M}$, and local spontaneous emission rates $\gamma_{S,M}^{\delta}$ indicate the change in the collective spin. The rate equations read
\begin{equation}
    \begin{split}
        \dot{p}_{S,M} &= -\big[\Gamma_{S,M}+ \xi^+_{S,M} + \xi^{-}_{S,M}\\
        &\qquad+\gamma^0_{S,M} + \gamma^+_{S,M} + \gamma^-_{S,M} \big]p_{S,M}\\ 
        &+\Gamma_{S,M+1}p_{S,M+1} + \gamma^{0}_{S,M+1}p_{S,M+1}\\
        &+\gamma^{+}_{S-1,M+1}p_{S-1,M+1} + \gamma^{-}_{S+1,M+1}p_{S+1,M+1}\\
        &+\xi^{+}_{S-1,M}p_{S-1,M} + \xi^{-}_{S+1,M}p_{S+1,M}.
    \end{split}
\end{equation}
The dynamics describe the evolution of the probability distribution $p_{S,M}$ on the Dicke triangle $\ket{S,M}$ where dephasing results in diffusion of the collective spin $S$, spontaneous emission results in spin diffusion and the simultaneous drift of the inversion $M$, and collective emission only describes the drift of the inversion.

\section{\texorpdfstring{Derivation of large $N$ expansion}{Derivation of large N expansion}}\label{app:large_N}

Using $\mathfrak{s}=S/N$ and $\mathfrak{m} = M/N$ as in the main text, the rates become, to lowest order in $1/N$

\begin{equation}
    \begin{aligned}
    \Gamma_{S,M} &\approx N^2\Gamma (\mathfrak{s}^2-\mathfrak{m}^2),\\
    \xi^{\pm}_{S,M} &\approx \frac{N\xi}{8\mathfrak{s}^2}(1\mp 2\mathfrak{s})(\mathfrak{s}^2-\mathfrak{m}^2), \\
    \gamma_{S,M}^{\pm}&\approx\frac{N\gamma}{8 \mathfrak{s}^2}(1\mp 2 \mathfrak{s}) (\mathfrak{s}\mp\mathfrak{m})^2,\\
    \gamma_{S,M}^{0}&\approx \frac{N\gamma}{4 \mathfrak{s}^2} (\mathfrak{s}-\mathfrak{m}) (\mathfrak{s}+\mathfrak{m}).
    \end{aligned}
\end{equation}

Thus, the mean-field equations of motion for $N\to\infty$ to leading order in $1/N$ are
\begin{equation}
\begin{aligned}
    \dot{\mathfrak{s}}&=\frac{1}{N}\left(\xi^+_{S,M}-\xi^-_{S,M}+\gamma^+_{S,M}-\gamma^-_{S,M}\right),\\
\dot{\mathfrak{m}}&=-\frac{1}{N}\left(\Gamma_{S,M}+\gamma^+_{S,M}+\gamma^0_{S,M}+\gamma^-_{S,M}\right).
\end{aligned}
\end{equation}
for which the expansion of our rates may be included to give 
\begin{equation}
\begin{aligned}
    \dot{\mathfrak{s}}&=-\frac{\xi}{2\mathfrak{s}}(\mathfrak{s}^2-\mathfrak{m}^2)-\frac{\gamma}{2\mathfrak{s}}(\mathfrak{s}^2+\mathfrak{m}^2+\mathfrak{m}),\\
    \dot{\mathfrak{m}}&=-N\Gamma(\mathfrak{s}^2-\mathfrak{m}^2)-\gamma\left(\mathfrak{m}+\frac{1}{2}\right).
\end{aligned}
\end{equation}

Defining the normalized intensity $\mathfrak I = \mathfrak{s}^2-\mathfrak{m}^2$ then yields the equations in the main text.

\section{Finite-size initial condition}
\label{app:mean_field_justification}

The hydrodynamic expression \(\mathfrak I\approx\mathfrak s^2-\mathfrak m^2\) vanishes at the fully inverted state because it neglects finite-size corrections. The exact finite-\(N\) normalized collective intensity of a Dicke state is
\begin{equation}
    \mathfrak I_N(S,M)
    =
    \frac{S(S+1)-M(M-1)}{N^2}.
\end{equation}
For the fully inverted initial state \(S=M=N/2\), this gives
\begin{equation}
    \mathfrak I_N\left(\frac{N}{2},\frac{N}{2}\right)
    =
    \frac{1}{N}.
\end{equation}
We therefore initialize the deterministic large-\(N\) equations with \(\mathfrak I(0)=1/N\). This is not a phenomenological seed, but the exact initial collective intensity retained as a finite-size correction to the hydrodynamic approximation. The resulting scaling exponents are compared with stochastic simulations in Fig.~\ref{fig:figS1}, with further finite-size analyses presented in Apps.~\ref{app:finite_size} and \ref{app:finite_size_spontaneous}.

\section{Analytical solutions to mean-field equations}\label{App:sect-analytical-mf}
In the most general case, we consider the nonlinear mean-field equations in the main text
\[
\begin{aligned}
    \partial_\tau\mathfrak{I} &= \left( 2\mathfrak{m} - g_\xi-g_\gamma\right) \mathfrak{I},\\
    \partial_\tau\mathfrak{m} &= -\mathfrak{I}-g_\gamma\left(\mathfrak{m}+\tfrac{1}{2}\right).
\end{aligned}
\]
These equations do not have an analytical solution in general. Nevertheless, it is possible to gain analytical insights. Without local decoherence $g_\xi = g_\gamma =0$ the equations reduce to pure collective decay, admitting an analytical solution. For $g_\gamma=0, g_\xi \neq 0$ an analytical solution remains possible, but not once $g_\gamma \neq 0$.
\subsection{Collective decay}
For purely collective decay ($g_\xi = g_\gamma = 0$) the equations reduce to 
\[
\begin{aligned}
    \partial_\tau\mathfrak{I} &= 2\mathfrak{m}\mathfrak{I},\\
    \partial_\tau\mathfrak{m} &= -\mathfrak{I}.
\end{aligned}
\]
By utilizing spin conservation \(\mathfrak I+\mathfrak m^2=\mathfrak s^2=\mathrm{const.}\), the dynamics can be reduced to a single differential equation for the inversion. The same invariant follows directly from the equations of motion, since
\[
    \frac{\mathrm d\mathfrak I}{\mathrm d\mathfrak m}=-2\mathfrak m\;\rightarrow\;\mathfrak I+\mathfrak m^2=\mathrm{const.}
\]
For the fully inverted state, \(\mathfrak s=1/2\). Within the hydrodynamic equations, retaining the exact finite-size seed \(\mathfrak I(0)=1/N\) therefore corresponds to
\[
\mathfrak m(0)=\sqrt{\mathfrak s^2-\frac{1}{N}}\approx\frac{1}{2}.
\]

Thus, 
\begin{equation}
\partial_\tau\mathfrak{m} = \mathfrak{m}^2 - \mathfrak{s}^2,
\end{equation}
which is the usual mean-field equation for the excitation number in Dicke superradiance. This differential equation, known as the logistic differential equation, is separable, integrable, and ultimately solvable for $\mathfrak{m}(\tau)$. We do not give the full calculation here, but we provide the solutions for completeness. Starting from full initial excitation $\mathfrak{m}(0) = 1/2$ and utilizing spin conservation $\mathfrak{s} = 1/2$ we obtain
\[
\begin{aligned}
    \mathfrak{I}(\tau) &=\mathfrak{s}^2\left\{1-\tanh^2\left[\mathfrak{s}(\tau - \tau_\star)\right]\right\},\\
    \mathfrak{m}(\tau) &= -\mathfrak{s}\tanh\left[\mathfrak{s}(\tau - \tau_\star)\right],
\end{aligned}
\]
with $\tau_\star = \ln(N)$.
In the original physical units, this corresponds to
\[
\begin{aligned}
    I(t) &= I_\star \left\{1 - \tanh^2\left[\frac{N\Gamma\left(t - t_\star\right)}{2}\right]\right\},\\
    M(t) & = -\frac{N}{2}\tanh\left[\frac{N\Gamma\left(t - t_\star\right)}{2}\right],
\end{aligned}
\]
with peak time $t_\star = \ln(N)/(N\Gamma)$ and peak intensity $I_\star = \Gamma N^2/4$.

\subsection{Collective decay with dephasing}

For collective decay with dephasing, the equations reduce to 
\[
\begin{aligned}
    \partial_\tau\mathfrak{I} &= \left( 2\mathfrak{m} - g_\xi\right) \mathfrak{I},\\
    \partial_\tau\mathfrak{m} &= -\mathfrak{I}.
\end{aligned}
\]
In this form, it is not immediately clear that these differential equations can still be solved analytically. However, defining $\tilde{\mathfrak{m}} = \mathfrak{m} -g_\xi/2$, the equations reduce to the previous case. Thus, reusing the same idea
\[
\frac{\text{d}\mathfrak{I}}{\text{d}\tilde{\mathfrak{m}}} = -2\tilde{\mathfrak{m}} \;\rightarrow\; \mathfrak{I} + \tilde{\mathfrak{m}}^2 = \text{const.}
\]
Thus, reusing the previous solution, we obtain for large $N$
\[
\begin{aligned}
    \mathfrak{I}(\tau) &=\frac{(1 - g_\xi)^2}{4}\left\{1 - \tanh^2\left[\frac{(1-g_\xi)}{2}\left(\tau - \tau_\star\right)\right]\right\},\\
    \mathfrak{m}(\tau) &= \frac{g_\xi}{2}-\frac{1-g_\xi}{2}\tanh\left[\frac{(1-g_\xi)}{2}\left(\tau - \tau_\star\right)\right],
\end{aligned}
\]
with $\tau_\star = \ln(N)/(1-g_\xi)$, which is simply a shifted and rescaled version of the previous solution. The renormalized burst time $\tau_\star$ is an approximation based on a large-$N$ expansion of the exact expression.

In the original physical units, the solution is
\[
\begin{aligned}
    I(t) &= I_\star \left\{1 - \tanh^2\left[\frac{N\Gamma(1 - g_\xi)}{2}\left(t - t_\star\right)\right]\right\},\\
    M(t) & = \frac{Ng_\xi}{2}-\frac{N(1-g_\xi)}{2}\tanh\left[\frac{N\Gamma(1-g_\xi)}{2}\left(t -t_\star\right)\right].
\end{aligned}
\]
with burst time $t_\star = \ln(N)/[N\Gamma(1-g_\xi)]$ and peak intensity $I_\star = \Gamma N^2(1-g_\xi)^2/4$.

\begin{figure*}[!tb]
    \centering
    \includegraphics[width=\linewidth]{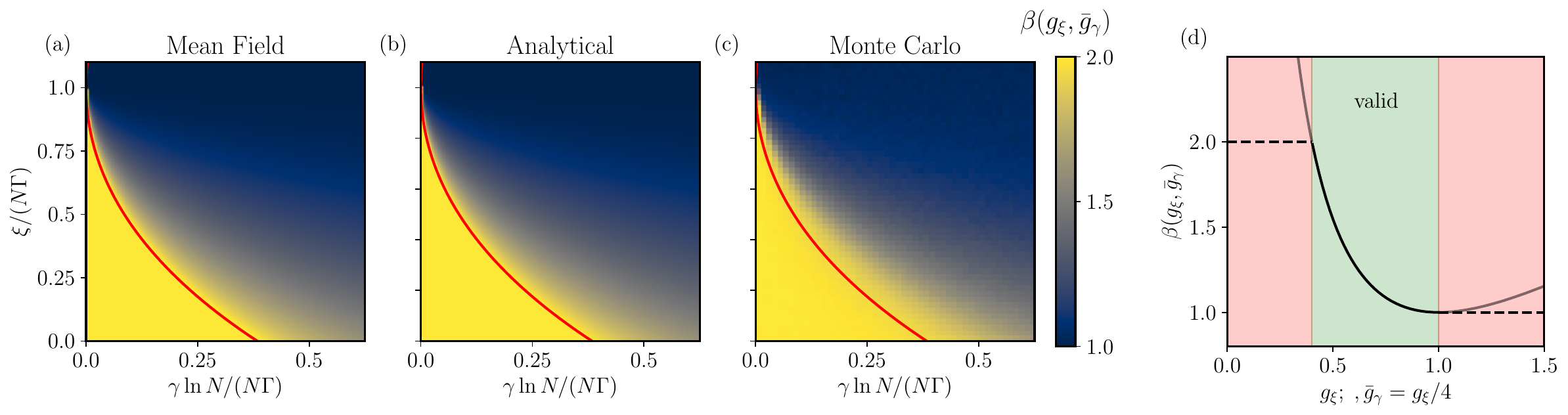}%
\caption{\textbf{Comparison of the scaling exponent} \(\beta(g_\xi,\bar g_\gamma)\). Each plotted point in (a) and (c) is an exponent extracted from a fit of the peak intensity over multiple system sizes to \(I_\star=A N^\beta\). (a) uses numerically integrated mean-field peak intensities over \(10^{10}\leq N\leq10^{23}\), (b) shows the analytical boundary-layer expression, and (c) uses Monte Carlo results using adaptive tau-leaping over \(3.3\times10^4\lesssim N\lesssim5.4\times10^8\). The agreement confirms the continuously varying subquadratic scaling \(I_\star\propto N^\beta\) in the amplification-limited regime. The red line marks the analytically derived boundary \(\beta=2\). (d) Evaluation of the analytical expression for the exponent $\beta$ from Eq.~\eqref{eq:beta_general} for the example in Fig.~\ref{fig:fig1}(b) in the main text. The analytical expression only holds in the green shaded part and not in the red shaded part, so only in the interval where $\beta$ is monotonic and capped by $1\leq \beta \leq 2$. The dashed black lines are added manually and are the physically correct value of the exponent in this regime.}
\label{fig:figS1}
\end{figure*}

\section{Asymptotic mean-field solution including spontaneous emission}\label{app:asymptotic_mean}

Consider the mean-field equations of motion 

\begin{equation}
\begin{aligned}
    \partial_\tau\mathfrak{I} &= \left( 2\mathfrak{m} - g_\xi-g_\gamma\right) \mathfrak{I},\\
    \partial_\tau\mathfrak{m} &= -\mathfrak{I}-g_\gamma\left(\mathfrak{m}+\tfrac{1}{2}\right).
\end{aligned}
\end{equation}

Based on the natural scaling $\bar g_\gamma=\frac{\gamma\ln N}{\Gamma N}$ so that $\lim_{N\to\infty} g_\gamma=0$, it becomes clear that asymptotically, this equation splits into two regions.

\begin{itemize}
    \item The linearized region, where \(\mathfrak I(\tau)\ll g_\gamma(\mathfrak m(\tau)+1/2)\) and we can approximate

\[
\begin{aligned}
    \partial_\tau\mathfrak{I} &= \left( 2\mathfrak{m} - g_\xi-g_\gamma\right) \mathfrak{I},\\
    \partial_\tau\mathfrak{m} &\approx -g_\gamma\left(\mathfrak{m}+\tfrac{1}{2}\right).
\end{aligned}
\]

\item The bulk region where \(\mathfrak I(\tau)\gg g_\gamma(\mathfrak m(\tau)+1/2)\), where $g_\gamma$ is negligible and

\[
\begin{aligned}
    \partial_\tau\mathfrak{I} &= \left( 2\mathfrak{m} - g_\xi\right) \mathfrak{I},\\
    \partial_\tau\mathfrak{m} &= -\mathfrak{I}.
\end{aligned}
\]

\end{itemize}

Both systems of equations are solvable, and the asymptotic scaling analysis is now simply a matter of stitching the solution of the linearized region onto the bulk region. Starting at $\mathfrak{m}(0)=1/2$ and $\mathfrak{I}(0)=1/N$ the solution in the linearized region becomes

\begin{equation}
\begin{aligned}\label{eq:linearized_solution}
    \mathfrak I(\tau)&= \frac{1}{N}\exp\left[\frac{2}{g_\gamma}\left(1-\me^{-g_\gamma\tau}\right)-(1+g_\gamma+g_\xi)\tau\right],\\
    \mathfrak{m}(\tau) &=\me^{-g_\gamma \tau}-\frac{1}{2}.
\end{aligned}
\end{equation}

\subsection{Phase boundary derivation}

If $\mathfrak I(\tau)$ reaches a maximal value proportional to $N^{\beta-2}$ with $1<\beta<2$ the maximum is reached when the linearization is still valid. The limiting cases $\beta\to 2$ and $\beta\to 1$, which determine the two phase boundaries, can thus also be determined from Eq.~\eqref{eq:linearized_solution}. We thus require that $\partial_\tau \mathfrak{I}(\tau_\star)=0$
\begin{equation}
\begin{aligned}
    \tau_\star&=\frac{1}{g_\gamma}\ln\left(\frac{2}{1+g_\xi+g_\gamma}\right),\\
    2\mathfrak m(\tau_\star)&=g_\xi+g_\gamma.
\end{aligned}
\end{equation}
and the intensity at this maximum yields
\begin{equation}
\begin{aligned}
    \ln \mathfrak I^{\text{lin}}_\star&=-\ln N+
    \frac{1-g_\xi-g_\gamma}{g_\gamma}\\
    &-\frac{1+g_\xi+g_\gamma}{g_\gamma}
    \ln\left(\frac{2}{1+g_\xi+g_\gamma}\right).
\end{aligned}
\end{equation}

Using \(\bar g_\gamma = g_\gamma \ln N\), this may be written as
\begin{equation}
    \mathfrak I_\star^{\mathrm{lin}}
    =
    N^{\beta(g_\xi,g_\gamma,\bar g_\gamma)-2},
\end{equation}
with
\begin{equation}
\begin{aligned}
    \beta(g_\xi,g_\gamma,\bar g_\gamma)
    &=1+\frac{1-g_\xi-g_\gamma}{{\bar g_\gamma}}\\
    &-\frac{(1+g_\xi+g_\gamma)\ln\left(\frac{2}{1+g_\xi+g_\gamma}\right)
    }{\bar g_\gamma}.
\end{aligned}
\end{equation}

The solution simplifies since $g_\gamma\to 0$ as $N\to\infty$ and hence
\begin{equation}
\label{eq:beta_general_app}
\begin{aligned}
    \beta(g_\xi,\bar g_\gamma)&=1+\frac{1-g_\xi-(1+g_\xi)\ln\left(\frac{2}{1+g_\xi}\right)}{\bar g_\gamma},
\end{aligned}
\end{equation}
recovering the expression from the main text. It applies only in the partially collective regime and is restricted to \(1\leq\beta\leq2\). The limiting values \(\beta\to2\) and \(\beta\to1\) define the boundaries beyond which the peak intensity exhibits quadratic and linear scaling, respectively [see Fig.~\ref{fig:figS1}(d)].

The phase boundary to the fully collective regime is obtained by letting $\beta\to 2$, yielding
\begin{equation}\label{eq:phase_boundary_app}
(1-g_\xi)-(1+g_\xi)\ln\left(\frac{2}{1+g_\xi}\right)=\bar g_\gamma.
\end{equation}

Letting $\beta\to 1$ leads to the phase boundary to the independent-emitter regime
\begin{equation}
    1-g_\xi-g_\gamma
    -
    (1+g_\xi+g_\gamma)\ln\left(\frac{2}{1+g_\xi+g_\gamma}\right)=0,
\end{equation}
which has the unique solution
\begin{equation}
    g_\xi+g_\gamma=1,
\end{equation}
where we kept $g_\gamma$ for completeness. The asymptotic condition is $g_\xi=1$ as in the pure dephasing case. This is the geometric condition that corresponds to a trajectory with an initial angle pointing into the bulk of the Dicke triangle rather than along the boundary. 

The resulting scaling exponent is compared in Fig.~\ref{fig:figS1} with exponents extracted from numerical data. Panel~(a) shows fits of the mean-field peak intensity over a wide range of system sizes to \(I_\star=A N^\beta\), while panel~(b) shows the analytical expression in Eq.~\eqref{eq:beta_general_app}. The mean-field results are essentially indistinguishable from the analytical prediction throughout the phase diagram. Panel~(c) shows the corresponding exponents extracted from Monte Carlo simulations. These reproduce the same overall scaling structure, but exhibit a softened crossover near the boundary between the fully and partially collective regimes, as expected from finite-size effects and the logarithmically slow convergence associated with \(g_\gamma=\bar g_\gamma/\ln N\). The analytical expression applies only in the partially collective regime \(1<\beta<2\); outside this interval, the physical exponent is fixed at \(\beta=2\) in the fully collective regime and at \(\beta=1\) in the independent-emitter regime, as illustrated for a representative cut in Fig.~\ref{fig:figS1}(d).

\begin{figure*}[!tb]
    \centering
    \includegraphics[width = \linewidth]{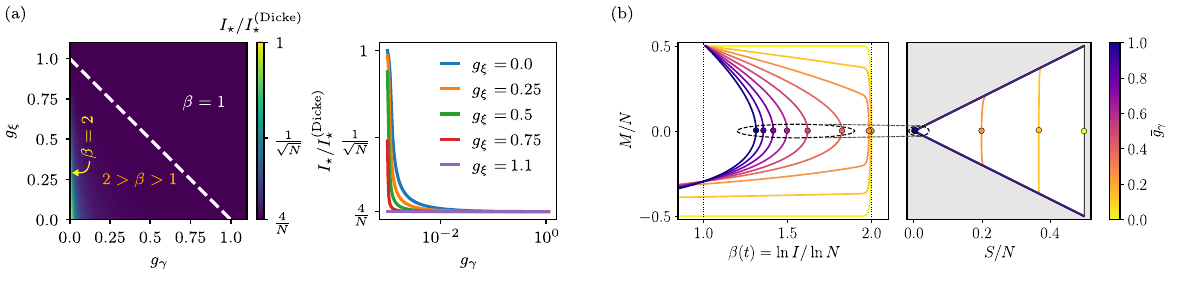}
    \caption{\textbf{Alternative phase diagram and scaling of trajectories in the Dicke triangle.} (a) Using $g_\gamma$ as the scaling parameter rather than $\bar g_\gamma$ we obtain an alternative phase diagram. Dashed white line gives the $\beta = 1$ boundary, while the $\beta =2$ boundary occurs now at $g_\gamma = 0$. We also show some selected cuts at different $g_\xi$ values. (b) On the right we show the Dicke triangle on a linear scale and on the left the magnetization as a function of the exponent $\beta$. For large $N$ the linear scaling on the Dicke triangle causes all trajectories with $\beta < 2$ to collapse to the outer boundary.}
    \label{fig:figS3}
\end{figure*}
\subsection{Intensity in the fully collective regime}

We now need to define a matching scale so that
\begin{equation}
    \lim_{N\to\infty}\mathfrak I_{\text{match}}=0,\qquad  \lim_{N\to\infty}\frac{\mathfrak I_{\text{match}}}{g_\gamma}=\infty,
\end{equation}
so that asymptotically $\mathfrak I(\tau)>\mathfrak I_{\text{match}}$ obeys the bulk equation. Since \(N^{\beta-2}\ll(\ln N)^{-1/2}\) for every fixed \(\beta<2\), trajectories with subquadratic peak intensity never reach the matching scale, whereas trajectories with \(I_\star\propto N^2\) do. One convenient choice for this matching scale is given by
\begin{equation}
    \mathfrak I_{\text{match}}=\frac{1}{\sqrt{\ln N}}.
\end{equation}

During the bulk evolution, spontaneous emission becomes parametrically irrelevant and the quantity
\begin{equation}
    \mathcal C=\mathfrak I+\left(\mathfrak m-\frac{g_\xi}{2}\right)^2,
\end{equation}
is conserved. Suppose that some fraction of excitations were drained during the linearized regime, yielding $\mathfrak m_\text{match}$. Then the bulk peak intensity becomes
\begin{equation}
    \mathfrak I_\star=\mathfrak I_{\text{match}}+\left(\mathfrak m_{\text{match}}-\frac{g_\xi}{2}\right)^2\xrightarrow{N\to\infty}\left(\mathfrak m_{\text{match}}-\frac{g_\xi}{2}\right)^2,
\end{equation}
which leads to $I_\star\propto N^2$ whenever the matching scale is reached during time evolution. The vanishing of its prefactor as it approaches the phase boundary is derived below.

\subsection{Analytical closing at the phase boundary}

To determine how the intensive peak intensity approaches zero for the fully collective to partially collective phase transition, we expand the intensity around a point $(g_\xi^b,\bar g_\gamma^b)$ such that $\beta(g_\xi^b,\bar g_\gamma^b)=2$. We thus expand the intensity in
\begin{equation}
\begin{aligned}
    \delta_\gamma=\bar g_\gamma^b-\bar g_\gamma,\qquad\delta_\xi=g_\xi^b- g_\xi,
\end{aligned}
\end{equation}
which also naturally leads to a change in the drained excitations $\delta_m=\mathfrak m-\mathfrak m_b$ where $\mathfrak m_b=g^b_\xi/2$. The peak intensity in terms of these becomes
\begin{equation}
    \mathfrak I_\star=\left(\delta_m+\frac{\delta_\xi}{2}\right)^2,
\end{equation}
and the phase boundary condition
\begin{equation}
\bar g_\gamma=1-2\mathfrak m+(1+g_\xi)\ln\left(\mathfrak m+\frac{1}{2}\right),
\end{equation}
expands as
\begin{equation}
    \frac{2}{1+g_\xi^b}\delta_m^2=\delta_\gamma+\ln\left(\frac{2}{1+g_\xi^b}\right)\delta_\xi,
\end{equation}
and the lowest-order contribution to the intensity thus yields
\begin{equation}
    \mathfrak I_\star=\frac{1+g_\xi^b}{2}\left(\delta_\gamma+\ln\left(\frac{2}{1+g_\xi^b}\right)\delta_\xi\right).
\end{equation}

This indicates that the intensity closes linearly in all directions transverse to the phase boundary for all $\bar g_\gamma>0$.

\section{Alternative phase diagram}\label{app:alternative_phasediagram}

In Sec.~\ref{app:asymptotic_mean} we chose to use $\bar g_\gamma$ as the scaling variable for spontaneous emission. This choice naturally emerges if the $N^2$ scaling should survive for a finite spontaneous emission rate. Using $\bar g_\gamma$ as the control parameter, finite spontaneous emission never results in independent emission. However, we also identified, from Eqs.~\eqref{eq:mf_complete}, that the criterion for amplification of the initial intensity seed is $g_\xi + g_\gamma < 1$. Thus, whenever \(g_\xi+g_\gamma<1\), the initial intensity seed is amplified. However, if $g_\gamma$ is chosen as the control parameter, the phase diagram becomes rather sparse [see Fig.~\ref{fig:figS3}(a)].

\section{Scaling structure of trajectories in the partially collective regime}\label{app:dicke_triangle_scaling}

For large $N$ and when including spontaneous emission, the Dicke triangle with linear axis scaling ceases to distinguish between trajectories with different scaling exponents $N^\beta$ where $\beta<2$. This occurs because in the thermodynamic limit all trajectories with $\beta < 2$ are along an infinitesimal layer along the $S=|M|$ boundary of the Dicke triangle. Or equivalently, all trajectories which enter the bulk of the Dicke triangle have $S\propto N$ and thus $I_\star \propto N^2$. In order to resolve this, we show in Fig.~\ref{fig:figS3}(b) an alternative representation of the trajectories, which allows us to distinguish trajectories with different $\beta$ values. However, this representation cannot distinguish between trajectories with different prefactors $I_\star = \alpha N^2$ and is less intuitive, so we do not use this representation in the main text.

Roughly, the scaling regimes and the corresponding trajectories can be classified in the following manner
\begin{itemize}[itemsep=2pt]
    \item $S\propto N \iff \beta = 2$: fully collective states with emission rates of order $N^2$
    \item $\sqrt{N} \ll S\ll N \iff 1 < \beta < 2$ for $M$ not of order $S$: partially collective states with superlinear but subquadratic emission rates.
    \item $S\left(S+1\right)\approx N/2+M^2 \iff \beta = 1$: the typical sector reached by fast dephasing, corresponding to independent-emitter dynamics.
    \item $S\left(S+1\right)<N/2+M^2 \iff \beta < 1$: subradiant relative to the independent-emitter rate on the same $M$ slice.
\end{itemize}

\begin{figure*}[t!]
    \centering
    \includegraphics[width=\linewidth]{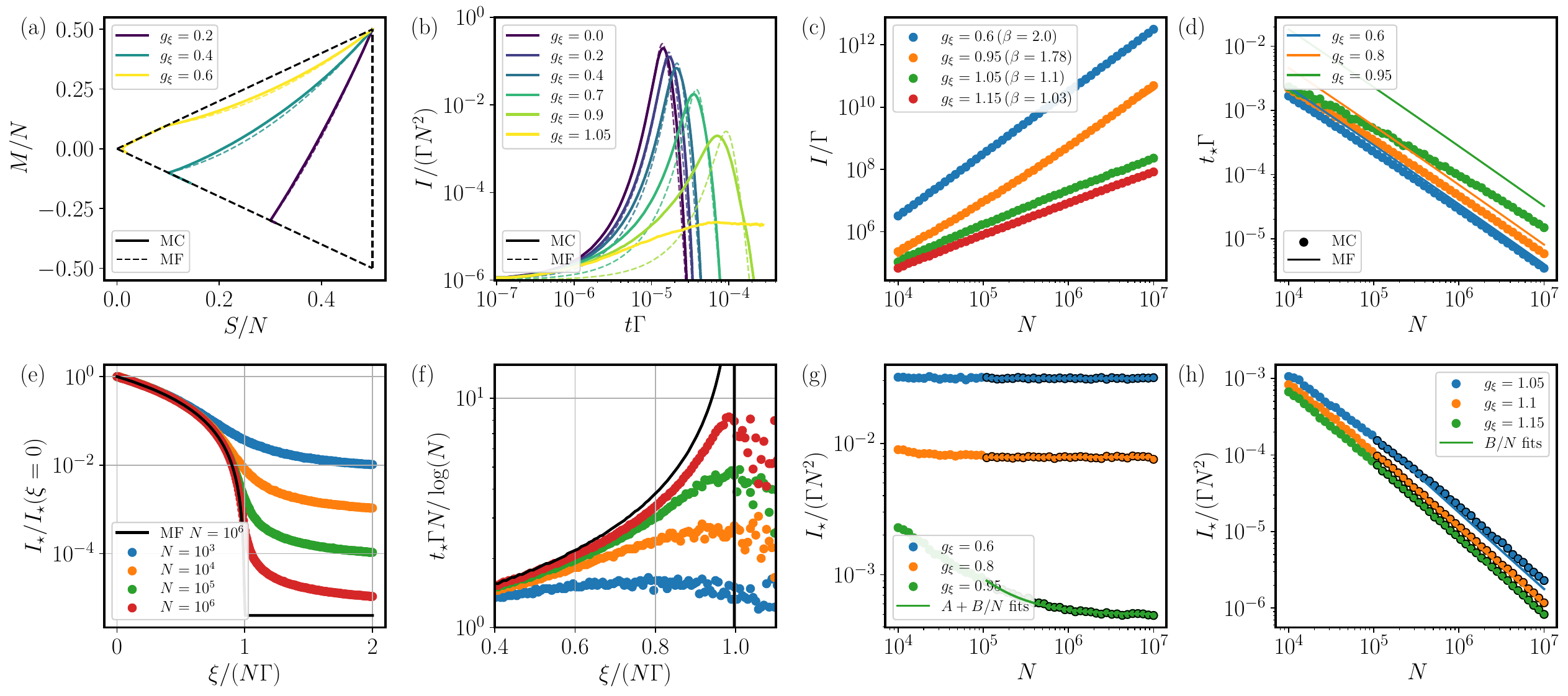}
    \caption{\textbf{Collective emission with local dephasing.}
    (a) Monte Carlo mean (solid), abbreviated as MC, compared to mean-field (dashed), abbreviated as MF, in the Dicke plane $(S/N, M/N)$ for increasing $g_\xi$ at fixed $\bar g_\gamma=0$.
(b) Intensity trajectories $I(t)/N^2$ for fixed $N=3\times 10^4$ and several $g_\xi$, comparing Monte Carlo (solid) and mean field (dashed). Mean field correctly captures the presence or absence of amplification.
(c) Scaling of $I_{\star}$ with system size $N$ (log--log) for representative $g_\xi$. The effective exponent changes from $\beta \approx 2$ below threshold to $\beta \approx 1$ above.
(d) Peak time $t_\star$ versus $N$, compared to the mean-field prediction $t_\star \sim \ln N/[N(1-g_\xi)]$. Convergence is rapid below threshold and slows near $g_\xi \simeq 1$.
(e) Maximal intensity $I_{\star}/N^2$ versus $g_\xi$ for several $N$. Data approach the mean-field prediction $I_{\star}/N^2 \to (1-g_\xi)^2/4$ below threshold and vanish above.
(f) Peak time $t_\star$ versus $g_\xi$, compared to the mean-field scaling $t_\star \sim \ln N/[N(1-g_\xi)]$. Agreement is good away from the threshold, with visible finite-size deviations near $g_\xi \simeq 1$.
(g) Below threshold: $I_{\star}/N^2$ versus $N$ with asymptotic fit $A + B/N$, demonstrating convergence to a finite value.
(h) Above threshold: $I_{\star}/N^2$ versus $N$ with fit $B/N$, consistent with the absence of a finite asymptotic contribution.}
    \label{fig:figS5}
\end{figure*}
\begin{figure*}[!thbp]
    \centering
    \includegraphics[width=\linewidth]{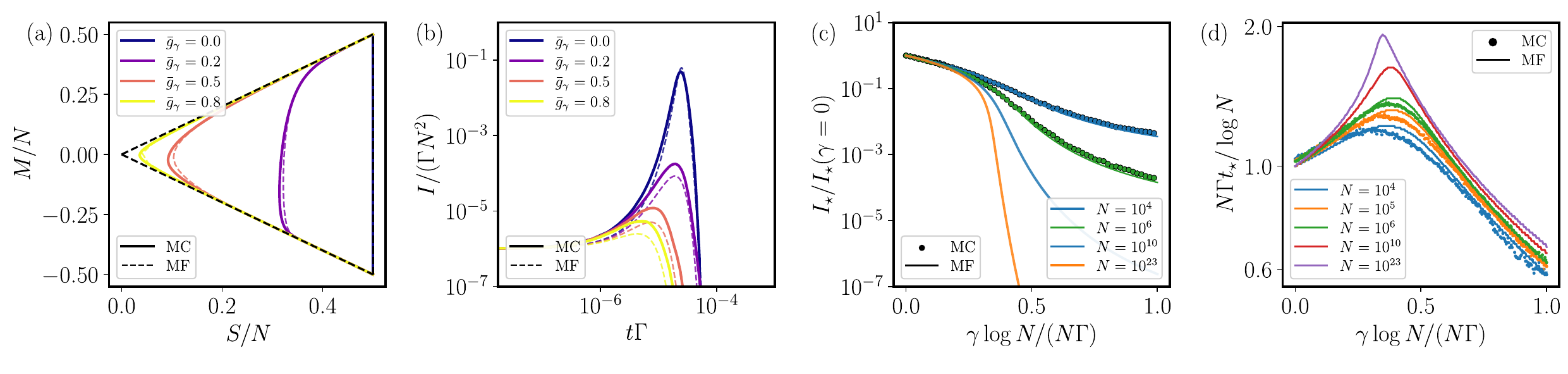}
    \caption{\textbf{Collective emission with local spontaneous emission.} (a) Comparison of Monte Carlo mean (solid) and mean field (dashed) in the Dicke triangle $(S/N,M/N)$ at fixed $g_\xi=0$ for increasing spontaneous-emission strength.
(b) Intensity trajectories $I(t)/N^2$ for the mean-field ($N\to\infty$) solution versus time for several values of the rescaled spontaneous-emission strength $\bar g_\gamma = \gamma \ln N/(N\Gamma)$, comparing Monte Carlo (solid) and mean field (dashed). Increasing $\bar g_\gamma$ progressively suppresses the superradiant burst.
(c) Maximal intensity $I_\star(\bar g_\gamma)$ normalized to its value at $\bar g_\gamma=0$, comparing Monte Carlo (markers) and mean-field (lines) results for moderate system sizes, together with mean-field curves at very large $N$.
(d) Rescaled peak time $N t_\star / \ln N$ as a function of $\bar g_\gamma$. A cusp develops only at extremely large $N$, indicating logarithmically slow convergence to the asymptotic limit.}
    \label{fig:figS6}
\end{figure*}
\section{Finite-size scaling analysis for dephasing}\label{app:finite_size}

We compare the mean-field treatment from the main text with exact stochastic simulations of the permutation-invariant rate equations (Eq.~\eqref{eq:dephasing_rates}) using Gillespie Monte Carlo. As shown in Fig.~\ref{fig:figS5}(a), the mean-field trajectories agree well with the Monte Carlo average throughout the bulk of the Dicke triangle. At the boundary, mean-field trajectories become stationary, while Monte Carlo trajectories continue to evolve due to boundary fluctuations absent in the mean-field. However, this affects only the long-time dynamics and does not influence the peak intensity, since trajectories reaching the boundary have already exited the amplifying region in the bulk of the Dicke triangle and cannot re-enter it.

Figure~\ref{fig:figS5}(b) shows intensity time traces for several values of $g_\xi$. Agreement is good away from the phase boundary and deteriorates near $g_\xi \simeq 1$, where the dynamics is increasingly dominated by fluctuations around the fully excited state. For $g_\xi > 1$, finite-$N$ trajectories can exhibit bursts. However, these do not scale as $N^2$ and vanish as $N\rightarrow \infty$.

This is confirmed in Fig.~\ref{fig:figS5}(c), where the maximal intensity approaches the mean-field prediction $\mathfrak I_\star = (1-g_\xi)^2/4$ below threshold and vanishes above it. Convergence becomes slower close to the critical point. The same trend is visible in the peak times shown in Fig.~\ref{fig:figS5}(d), which are more sensitive to fluctuations and therefore display stronger finite-size effects.

A more detailed finite-size scaling analysis is shown in Fig.~\ref{fig:figS5}(e-h). Below threshold, $I_{\star}/N^2$ converges to a finite value, while above threshold it vanishes as $1/N$, consistent with the absence of collective enhancement. The effective scaling exponent, therefore, changes from $\beta \approx 2$ below threshold to $\beta \approx 1$ above. Peak times exhibit slower convergence near the critical point, reflecting the increasing role of fluctuations in this regime.

\section{Finite-size scaling analysis for spontaneous emission}\label{app:finite_size_spontaneous}

We now switch the finite-size analysis to collective emission with local spontaneous emission (without dephasing). In contrast to the dephasing-only case, mean-field and Monte Carlo dynamics remain in good quantitative agreement across the full trajectory, since spontaneous emission provides a deterministic decay channel that removes the boundary artifact present in the dephasing-only mean field.

As shown in Fig.~\ref{fig:figS6}(a), trajectories remain well described by mean-field dynamics. Spontaneous emission suppresses the buildup of large-$S$ components and drives the system away from the cooperative edge of the Dicke triangle. The corresponding intensity traces in Fig.~\ref{fig:figS6}(b) show good qualitative agreement between Monte Carlo and mean field.

The nontrivial aspect of this regime lies in the asymptotic structure of the mean-field solution. Due to the logarithmic amplification time $\tau \propto \ln N$, convergence to the $N\to\infty$ limit is logarithmically slow. This is illustrated in Fig.~\ref{fig:figS6}(c), where the rescaled intensity has not fully converged to the asymptotic limit even for very large system sizes, but the mean-field agrees well with the Monte Carlo results.

A similar behavior is observed for the peak times in Fig.~\ref{fig:figS6}(d), where convergence to the asymptotic solution remains slow. The corresponding phase diagram in Fig.~\ref{fig:figS1}(c) agrees with the analytic boundary derived in the main text.

\bibliography{bibliography}
\end{document}